\newcommand{\roughly}[1]%
    {{\mathrel{\raise.3ex\hbox{$#1$\kern-.75em\lower1ex\hbox{$\sim$}}}}}

\documentclass[12pt]{article}
\usepackage{paper2e}
\usepackage{latexsym}
\usepackage{graphicx}

\begin{document}
\begin{titlepage}

\preprint{IPMU09-0066}

\title{Dark matter as integration constant in Ho\v{r}ava-Lifshitz
 gravity}

\author{Shinji Mukohyama}

\address{
Institute for the Physics and Mathematics of the Universe (IPMU)\\ 
The University of Tokyo\\
5-1-5 Kashiwanoha, Kashiwa, Chiba 277-8582, Japan
}

\begin{abstract}

 In the non-relativistic theory of gravitation recently proposed by
 Ho\v{r}ava, the Hamiltonian constraint is not a local equation
 satisfied at each spatial point but an equation integrated over a whole
 space. The global Hamiltonian constraint is less restrictive than its
 local version, and allows a richer set of solutions than in general
 relativity. We show that a component which behaves like pressureless
 dust emerges as an ``integration constant'' of dynamical
 equations and momentum constraint equations. Consequently, classical
 solutions to the infrared limit of Ho\v{r}ava-Lifshitz gravity can
 mimic general relativity plus cold dark matter. 

\end{abstract}

\end{titlepage}

\section{Introduction}
\label{sec:introduction}

Dark energy and dark matter are two major mysteries in modern
cosmology. Assuming that general relativity is correct at long distances
up to cosmological scales, precision observational data indicates that
more than 90\% of our universe consists of dark energy and dark 
matter. Although some gravitational properties of the dark components
are known, they are not optically observed and, thus, we do not know
what they really are. This situation makes us suspect that modifying
gravity in the infrared (IR) might address the mysteries of dark energy
and/or dark matter.

Recently a power-counting renormalizable~\footnote{
Note, however, that renormalizability has not yet been established in a
rigorous manner beyond the level of power-counting.} theory
of gravitation was proposed by
Ho\v{r}ava~\cite{Horava:2009uw,Horava:2008ih}. One of the most important 
aspects of the theory is that in the ultraviolet (UV) it is
fundamentally non-relativistic and exhibits the Lifshitz scale
invariance 
%
\begin{equation}
 t \to b^z t, \quad \vec{x}\to b \vec{x},
  \label{eqn:scaling}
\end{equation}
with dynamical critical exponent $z=3$. Ho\v{r}ava's theory is
considered as a potential candidate for the theory of quantum gravity
and is often called Ho\v{r}ava-Lifshitz gravity. Various aspects of this 
theory have been
investigated~\cite{Takahashi:2009wc}-\cite{Calcagni:2009qw}.

Ho\v{r}ava-Lifshitz gravity has not yet been intended to be a unified
theory. Clearly, further developments or/and embedding into a ``bigger''
theory is needed. For example, since the ``limit of speed'' is an
emergent quantity in the IR, different species including those in the
standard model of particle physics must be related to each other in the
framework of Ho\v{r}ava's theory so that the ``limits of speed'' for
different species in the IR agree with the ``velocity of
light''~\footnote{See
e.g. refs.~\cite{Coleman:1998ti,Jacobson:2001tu,Moore:2001bv} for tight
experimental limits on Lorentz violation.}. This obviously indicates
that embedding of this theory into a unified theory (or other way
around) is necessary for the theory to be a part of the real world.

Still, it is interesting to investigate universal properties of the
theory~\footnote{In this respect, the so called detailed balance
condition is neither essential nor universal for Ho\v{r}ava-Lifshitz
gravity as already stated in \cite{Horava:2009uw}.} and its cosmological
implications, in parallel with those fundamental issues. For example,
the $z=3$ Lifshitz scaling not only is the origin of the power-counting
renormalizability but also leads to a number of interesting cosmological
consequences, such as generation of scale-invariant cosmological
perturbations from a non-inflationary epoch of the early
universe~\cite{Mukohyama:2009gg} and a particular scaling of radiation
energy density ($\propto a^{-6}$)~\cite{Mukohyama:2009zs}.

The purpose of the present paper is to point out that 
Ho\v{r}ava-Lifshitz gravity can mimic general relativity plus cold dark
matter.

\section{Basic idea}
\label{sec:idea}

Before explaining why the IR limit of Ho\v{r}ava-Lifshitz gravity can
behave like general relativity plus cold dark matter, let us remind
ourselves about the structure of Einstein's general relativity since the 
existence of dark matter was suspected by assuming general
relativity. General relativity fully respects $4$-dimensional spacetime
diffeomorphism invariance as the fundamental symmetry of the theory. As
a result, it has four constraint equations: one called Hamiltonian
constraint and three called momentum constraint. These constraints must
be satisfied at each spatial point at each time. However, since the
constraint equations are preserved under time evolution by dynamical
equations, i.e. other components of the Einstein equation, it is also
possible to impose the constraint equations only on an initial
hypersurface and to solve dynamical equations afterwards. In this case,
constraint equations are automatically satisfied at late time.

As an illustration, let us consider a flat Friedmann-Robertson-Walker
(FRW) spacetime driven by components with equations of state
$P_i=P_i(\rho_i)$, where $\rho_i$ and $P_i$ are energy density and
pressure of the $i$-th component. Because of the spatial homogeneity,
the momentum constraint is trivial. On the other hand, the Hamiltonian
constraint gives the famous Friedmann equation: 
%
\begin{equation}
 3\frac{\dot{a}^2}{a^2} = 8\pi G_N \sum_{i=1}^n\rho_i,
  \label{eqn:Friedmann}
\end{equation}
where $a$ is the scale factor of the universe, a dot represents time
derivative and $n$ is the number of components. The conservation of
stress energy tensor states that 
%
\begin{equation}
 \dot{\rho}_i + 3\frac{\dot{a}}{a}(\rho_i+P_i) = 0.
  \label{eqn:conservation}
\end{equation}
These $n+1$ equations are sufficient to predict future evolution of the
universe, provided that the initial value of $a$ and $\rho_i$ are 
specified. The remaining non-trivial component of the Einstein equation
gives the dynamical equation
%
\begin{equation}
 -2\frac{\ddot{a}}{a}-\frac{\dot{a}^2}{a^2}
  = 8\pi G_N \sum_{i=1}^nP_i,
  \label{eqn:dynamical}
\end{equation}
but this follows from the previous $n+1$ equations. Therefore, it
suffices to solve the Friedmann equation (\ref{eqn:Friedmann}) coupled
with the conservation equation (\ref{eqn:conservation}). However, it 
is also consistent to solve the dynamical equation (\ref{eqn:dynamical})
coupled with the conservation equation (\ref{eqn:conservation}),
provided that the Friedmann equation (\ref{eqn:Friedmann}) is imposed at
an initial time. In other words, the Friedmann equation can be 
considered as an first integral of the dynamical equation with a special
choice of an integration constant.

Now, let us suppose that there is a theory without Hamiltonian
constraint. Let us, however, suppose that in the FRW spacetime, we still
have the conservation equation (\ref{eqn:conservation}) and the 
dynamical equation (\ref{eqn:dynamical}) . This is perfectly fine as we
have $n+1$ independent differential equations for $n+1$ variables,
$a(t)$ and $\rho_i(t)$ ($i=1,\cdots,n$). Actually, we obtain 
%
\begin{equation}
 3\frac{\dot{a}^2}{a^2} = 8\pi G_N 
  \left(\sum_{i=1}^n\rho_i+\frac{C}{a^3}\right)
  \label{eqn:modFriedmann}
\end{equation}
as a first integral of the dynamical equation, where $C$ is an
integration constant, and this is almost the same as the Friedmann
equation (\ref{eqn:Friedmann}). The only difference is the term
$C/a^3$. What is interesting is that this is exactly of the form of dark
matter. In general relativity, dark matter ($\propto a^{-3}$) is
included as one of $\rho_i$'s and, thus, must be derived from an action 
principle since $\rho_i$ is a component of a stress-energy tensor. In
general relativity this is the origin of the mystery: we need to explain
what dark matter is made of by specifying its action. On the other hand,
in this hypothetical theory without Hamiltonian constraint, the term
proportional to $a^{-3}$ emerges as an integration constant and, thus, 
we do not need an action for it.

Intriguingly enough, as we shall briefly explain in the next paragraph,
in Ho\v{r}ava-Lifshitz gravity there is no Hamiltonian constraint as a
local equation at each spatial point. Instead, the Hamiltonian
constraint equation in Ho\v{r}ava-Lifshitz gravity is an equation
integrated over a whole space. In homogeneous spacetime such as the FRW
spacetime, the global Hamiltonian constraint is as good as local one
since all spatial points are equivalent. However, in inhomogeneous
spacetimes there can be drastic differences. If the whole universe is
much larger than the present Hubble volume then it is possible that the
universe far beyond the present Hubble horizon is different from our
patch of the universe inside the horizon. In this case, the global
Hamiltonian constraint does not restrict the universe inside the
horizon. Even if we approximate our patch of the universe inside the
present horizon by the FRW spacetime, the whole universe can include
inhomogeneities of super-horizon scales and, thus, the global
Hamiltonian constraint does not restrict the FRW spacetime which just
approximates the behavior inside the horizon. Therefore, as in the
hypothetical theory considered in the previous paragraph, the absence of
local Hamiltonian constraint in Ho\v{r}ava-Lifshitz gravity results in
an extra term $\propto a^{-3}$ in the ``Friedmann equation'' or, to be
precise, the first integral of the dynamical equation. As before, this
term can mimic dark matter but we do not need an action for it.

Absence of local Hamiltonian constraint in Ho\v{r}ava-Lifshitz gravity
originates from the projectability of the lapse function. The basic
quantities in Ho\v{r}ava-Lifshitz gravity are the $3$-dimensional
spatial metric $g_{ij}$, the shift vector $N^i$ and the lapse function
$N$. In terms of these quantities the $4$-dimensional spacetime metric
is written in the ADM form: 
%
\begin{equation}
 ds^2 = -N^2dt^2 + g_{ij}(dx^i+N^idt)(dx^j+N^jdt). 
  \label{eqn:ADM}
\end{equation}
The former two, $g_{ij}$ and $N^i$, can depend on both spatial 
coordinates $x^k$ and the time variable $t$. On the other hand, the
projectability condition states that the lapse function $N$ should
depend only on $t$ and be independent of spatial coordinates. 
The projectability of the lapse function stems from the fundamental
symmetry of the theory, i.e. invariance under the foliation-preserving
diffeomorphism: 
%
\begin{equation}
 x^i \to \tilde{x}^i(x^j,t), \quad t \to \tilde{t}(t), 
\end{equation}
and therefore must be respected. Essentially, the lapse function $N$
represents a gauge degree of freedom associated with the
space-independent time reparametrization. Thus, it is very natural to
restrict $N$ to be independent of spatial coordinates. This is the
projectability condition. Of course, a space-independent $N$ cannot be
transformed to a space-dependent function by foliation-preserving
diffeomorphism. This point was already made clear by
Ho\v{r}ava~\cite{Horava:2009uw}~\footnote{In the last paragraph of
subsection 2.1 of \cite{Horava:2009uw}, it says that, except for the
case with extra symmetry such as the Weyl symmetry ($\lambda=1/3$),
fluctuations of the lapse function must be space-independent. 
\label{footnote:projectability}}.

\section{IR limit of Ho\v{r}ava-Lifshitz gravity}
\label{sec:IRlimit}

In the IR limit the action of Ho\v{r}ava-Lifshitz gravity is reduced to 
%
\begin{equation}
 I_{HL} = \frac{1}{16\pi G_N}\int dtd^3x \sqrt{g}N
  \left[ K_{ij}K^{ij}-\lambda K^2 + R - 2\Lambda \right],
\end{equation}
where $K_{ij}=(\dot{g}_{ij}-D_i N_j-D_j N_i)/(2N)$ is the extrinsic
curvature of the constant time hypersurface, $K=K^i_i$, and $D$ is the
$3$-dimensional covariant derivative compatible with $g_{ij}$. This
looks identical to the Einstein-Hilbert action in the ADM form if and
only if $\lambda=1$. Hence, hereafter, we assume that the
renormalization group (RG) flow brings $\lambda$ to $1$ in the IR or
that $\lambda$ stays at $1$ from higher energy scales all the way down 
to the IR under the RG flow. The RG flow of Ho\v{r}ava-Lifshitz gravity
has not been investigated in details and, thus, must be addressed in the
future. In this paper, we simply assume that $\lambda=1$ is an IR fixed 
point of the RG flow.

Even with $\lambda=1$, however, there is an important difference between 
the IR limit of Ho\v{r}ava-Lifshitz gravity and general relativity. In 
Ho\v{r}ava-Lifshitz gravity the projectability condition requires that 
the lapse function $N$ should depend only on $t$. Because of this
restriction, the Hamiltonian constraint, i.e. the equation derived from
functional derivative of the total action with respect to the lapse
function, is not a local equation but an equation integrated over a
constant time hypersurface: 
%
\begin{equation}
 \int d^3x\sqrt{g} (G^{(4)}_{\mu\nu}
  +\Lambda g^{(4)}_{\mu\nu}
  -8\pi G_N T_{\mu\nu})n^{\mu}n^{\nu} = 0. 
  \label{eqn:Hamiltonian-constraint}
\end{equation}
Here, $g^{(4)}_{\mu\nu}$ is the $4$-dimensional metric shown in
(\ref{eqn:ADM}), $G^{(4)}_{\mu\nu}$ is the corresponding $4$-dimensional
Einstein tensor, $T_{\mu\nu}$ is the stress energy  tensor, and
$n^{\mu}$ is the unit normal to the constant time hypersurface given by 
%
\begin{equation}
 n_{\mu}dx^{\mu} = -Ndt, \quad 
  n^{\mu}\partial_{\mu} = \frac{1}{N}(\partial_t-N^i\partial_i). 
  \label{eqn:unitnormal}
\end{equation}
On the other hand, the momentum constraint and the dynamical equations
are local equations as in general relativity: 
%
\begin{equation}
 (G^{(4)}_{i\mu}+\Lambda g^{(4)}_{i\mu}
  -8\pi G_N T_{i\mu})n^{\mu} = 0,
  \label{eqn:momentum-constraint}
\end{equation}
and 
%
\begin{equation}
 G^{(4)}_{ij}+\Lambda g^{(4)}_{ij}-8\pi G_N T_{ij} = 0.
  \label{eqn:dynamical-eq}
\end{equation}

The Hamiltonian constraint (\ref{eqn:Hamiltonian-constraint}) and
momentum constraint (\ref{eqn:momentum-constraint}) are preserved by the
dynamical equations (\ref{eqn:dynamical-eq}). Thus, it suffices to solve
the dynamical equations, provided that the initial condition satisfies
the constraint equations. Note that the global Hamiltonian constraint 
(\ref{eqn:Hamiltonian-constraint}) is less restrictive than its local
version, and allows a richer set of solutions than in general
relativity.

\section{Dark matter as ``integration constant''}
\label{sec:CDM}

Let us define deviation from general relativity $T^{HL}_{\mu\nu}$ by 
%
\begin{equation}
 T^{HL}_{\mu\nu} \equiv \frac{1}{8\pi G_N}
  \left(G^{(4)}_{\mu\nu}+\Lambda g^{(4)}_{\mu\nu}\right)
  - T_{\mu\nu},
\end{equation}
or 
%
\begin{equation}
 G^{(4)}_{\mu\nu}+\Lambda g^{(4)}_{\mu\nu}
  = 8\pi G_N \left(T_{\mu\nu} +  T^{HL}_{\mu\nu} \right). 
\end{equation}
This looks like Einstein equation with the dark sector
$T^{HL}_{\mu\nu}$. In the IR, not only the gravitational sector but also
the (real) matter sector should respect the full $4$-dimensional
diffeomorphism invariance. Thus, the conservation of energy momentum
tensor for real matter $\nabla^{\mu}T_{\mu\nu}=0$ should hold in 
the IR, where $\nabla$ is the $4$-dimensional covariant derivative
compatible with $g^{(4)}_{\mu\nu}$. The Bianchi identity then implies
the conservation $\nabla^{\mu}T^{HL}_{\mu\nu}=0$ of the dark sector. If
the $4$-dimensional diffeomorphism invariance is slightly broken in the
(real) matter sector then
$\nabla^{\mu}T^{HL}_{\mu\nu}=-\nabla^{\mu}T_{\mu\nu}\ne 0$. 
Note that invariance of the (real) matter action under $3$-dimensional
spatial diffeomorphism implies that $\nabla^{\mu}T_{\mu i}=0$ and thus
$\nabla^{\mu}T_{\mu\nu}\propto n_{\nu}$.

The field equations in the IR limit of Ho\v{r}ava-Lifshitz gravity is
now written in terms of $T^{HL}_{\mu\nu}$. The Hamiltonian constraint 
(\ref{eqn:Hamiltonian-constraint}) is 
%
\begin{equation}
 \int d^3x\sqrt{g} T^{HL}_{\mu\nu}n^{\mu}n^{\nu} = 0.  
\end{equation}
The momentum constraint (\ref{eqn:momentum-constraint}) and dynamical
equations (\ref{eqn:dynamical-eq}) are 
%
\begin{equation}
 T^{HL}_{i\mu}n^{\mu} = 0,
\end{equation}
and 
%
\begin{equation}
 T^{HL}_{ij} = 0.
\end{equation}
As a general solution to the momentum constraint and dynamical
equations, we obtain
%
\begin{equation}
 T^{HL}_{\mu\nu} = \rho^{HL} n_{\mu}n_{\nu}, 
\end{equation}
where $\rho^{HL}$ is a scalar function of spacetime coordinates
$(t,x^i)$. This is equivalent to the stress energy tensor of a
pressureless dust with energy density $\rho^{HL}$ and the unit tangent
$n^{\mu}$ to its flow. Note that $n^{\mu}$ is tangent to a congruence of
geodesics: 
%
\begin{equation}
 n^{\mu}\nabla_{\mu}n_{\nu} =
  n^{\mu}\nabla_{\nu}n_{\mu} =
  \frac{1}{2}\partial(n^{\mu}n_{\mu}) =
  0. \label{eqn:geodesic}
\end{equation}
Here, for the first equality, we have used the expression
(\ref{eqn:unitnormal}) and the fact that the lapse function $N$ depends
only on $t$. Finally, the Hamiltonian constraint is 
%
\begin{equation}
 \int d^3x\sqrt{g}\rho^{HL} = 0. 
  \label{eqn:totalenergy}
\end{equation}
This states that the total energy of the dust-like component in the dark
sector should vanish. Of course $\rho^{HL}$ can be positive everywhere
in our patch of the universe. As already explained in
Sec.~\ref{sec:idea}, this applies even when we approximate our universe
by a FRW spacetime since overall homogeneity inside the current Hubble
volume does not exclude super-horizon inhomogeneities.

As stated at the end of the previous section, the dynamical equations
preserve the constraint equations. Thus, it suffices to solve the
``modified Einstein equation'' 
%
\begin{equation}
 G^{(4)}_{\mu\nu}+\Lambda g^{(4)}_{\mu\nu}
  = 8\pi G_N
  \left(T_{\mu\nu} +  \rho^{HL} n_{\mu}n_{\nu} \right), 
  \label{eqn:modEin}
\end{equation}
coupled with field equations of real matter fields, provided that the
initial condition of the dark sector satisfies the global Hamiltonian
constraint (\ref{eqn:totalenergy}). Note that $\rho^{HL}$ does not have
to vanish everywhere. It can be positive somewhere in the universe and 
negative elsewhere, as far as it sums up to zero. For example,
$\rho^{HL}$ can be positive everywhere in our patch of the universe
inside the present Hubble horizon. Note also that the additional term
$\rho^{HL} n_{\mu}n_{\nu}$ is just an ``integration constant'' and does
not represent a real dust. In other words, this additional term acts as
cold dark matter but does not require an action.

The Bianchi identity applied to the ``modified Einstein equation''
(\ref{eqn:modEin}) leads to 
%
\begin{equation}
 \left(\partial_{\perp}\rho^{HL}+K\rho^{HL}\right)n_{\nu}
  = -\nabla^{\mu}T_{\mu\nu},
  \label{eqn:DMconservation-pre}
\end{equation}
where $\partial_{\perp}=n^{\mu}\partial_{\mu}$ and $\nabla$ is the
$4$-dimensional covariant derivative compatible with
$g^{(4)}_{\mu\nu}$. This is consistent with the fact that the
$3$-dimensional spatial diffeomorphism invariance of the (real) matter
action implies $\nabla^{\mu}T_{\mu\nu}\propto n_{\nu}$. Therefore, we
obtain 
%
\begin{equation}
 \partial_{\perp}\rho^{HL}+K\rho^{HL}
  =  n^{\nu}\nabla^{\mu}T_{\mu\nu}. 
\end{equation}
The right hand side acts as a source term for the ``dark matter'' energy
density $\rho^{HL}$, and is non-vanishing if the (real) matter sector
breaks a part of $4$-dimensional diffeomorphism invariance.

If the gravity sector also breaks the $4$-dimensional diffeomorphism,
i.e. if $\lambda$ slightly deviates from $1$ or/and higher spatial
curvature terms become non-negligible, then the left hand side of the
``modified Einstein equation'' (\ref{eqn:modEin}) gets corrections as
%
\begin{equation}
 G^{(4)}_{\mu\nu}+\Lambda g^{(4)}_{\mu\nu}
  + O(\lambda - 1) + (\mbox{higher curvature corrections})
  = 8\pi G_N
  \left(T_{\mu\nu} +  \rho^{HL} n_{\mu}n_{\nu} \right). 
  \label{eqn:modEin-general}
\end{equation}
The new terms in the left hand side result in extra contributions to 
the right hand side of (\ref{eqn:DMconservation-pre}) but they are again
proportional to $n_{\nu}$ due to $3$-dimensional spatial diffeomorphism
invariance. Therefore, in general we have 
%
\begin{equation}
 \partial_{\perp}\rho^{HL}+K\rho^{HL}
  =  n^{\nu}\nabla^{\mu}T_{\mu\nu} + O(\lambda -1)
  + (\mbox{higher curvature corrections}). 
  \label{eqn:DMconservation}
\end{equation}
The right hand side is non-vanishing and the ``dark matter'' is
inevitably ``generated'' if either the (real) matter sector or the
gravity sector breaks a part of $4$-dimensional diffeomorphism
invariance. Thus, turning off the ``dark matter'' completely or,
equivalently, imposing a local Hamiltonian constraint is not a
consistent truncation in general.

In the early universe, the r.h.s. of (\ref{eqn:DMconservation}) is
non-vanishing. Quantum fluctuations of scalar graviton, tensor graviton
and (real) matter fields act as source of ``dark matter''. Once a
cosmological model is specified, it is possible to predict the typical
amplitude of $\rho^{HL}$ generated in the early universe. (Such quantum
fluctuations include modes with various wavelengths, including those
much longer than the size of the present Hubble scale.) For this reason,
the initial condition of ``dark matter'' is not arbitrary and this
scenario has predictability.

On the other hand, at late time the r.h.s. of (\ref{eqn:DMconservation})
should vanish if matter sector recovers $4$-dimensional diffeomorphism
in the IR and if $\lambda=1$ is a stable IR fixed point of the RG
flow. In this case, (\ref{eqn:DMconservation}) is reduced to the usual
conservation law for ``dark matter'': 
%
\begin{equation}
 \partial_{\perp}\rho^{HL}+K\rho^{HL}  =  0.
  \label{eqn:DMconservationIR}
\end{equation}

The flow of ``dark matter'' is tangent to the vector $n^{\mu}$ defined
in (\ref{eqn:unitnormal}) and thus is orthogonal to the constant time
hypersurface. When a cusp is about to form, the spatial curvature of the
constant time hypersurface increases. The system enters the
non-relativistic regime and higher spatial curvature terms become
important. Among them, terms with $z=3$ generate the strongest
restoring force. Also, $\lambda$ may deviate from $1$ by RG flow. As in
some early universe
models~\cite{Calcagni:2009ar,Kiritsis:2009sh,Brandenberger:2009yt}, we
expect that the would-be cusp can bounce at short distance scales. Note
that this is not because of deviation from geodesics~\footnote{
In the case of ghost condensate~\cite{ArkaniHamed:2003uy}, the
derivative of the scalar field responsible for the condensate deviates
from geodesics because of higher derivative
terms~\cite{ArkaniHamed:2005gu}. On the other hand, in
Ho\v{r}ava-Lifshitz gravity the vector $n^{\mu}$ always satisfies the
geodesic equation (\ref{eqn:geodesic}).} but because of repulsive
gravity. In usual situation, congruence of geodesics would form caustics
and thus cusps because gravity is attractive. On the other hand, for the
flow of ``dark matter'' proposed in the present paper, higher curvature
terms become important near the would-be cusps and provide negative
effective energy and repulsive gravity. That is the reason why we expect
bounce. Note that the bounce is provided by nonlinear terms and thus
nonlinear analysis is needed. It is also important to include
backreactions of the higher spatial curvature terms to the geometry
since, as mentioned in the previous paragraph, the bounce is not due to
deviation from geodesics but due to repulsive gravity at short
distances. Without taking into account nonlinear terms and backreactions
to the geometry, we would never be able to describe ``dark matter''
properly.

The initial value formulation of Ho\v{r}ava-Lifshitz gravity consists of
dynamical equation, global Hamiltonian constraint, local momentum
constraint and gauge conditions. In this language, the ``dark matter''
emerges only after solving the system of equations when we try to
interpret a solution. Therefore, as far as scalar graviton, tensor
graviton and matter fields are properly included as dynamical variables,
we do not have to consider ``dark matter'' as an independent dynamical
field. This structure should persist even in quantum level. For this
reason, we do not have to promote the ``dark matter'' from non-dynamical
integration ``constant'' to a dynamical field. Nonetheless, when we
interprete a solution after solving the dynamics of the system either
classically or quantum mechanically, we see that difference from general
relativity amounts to ``dark matter''.

\section{Summary and discussion}
\label{sec:summary}

In the non-relativistic, power-counting renormalizable theory of
gravitation recently proposed by Ho\v{r}ava, the so called
projectability condition must be respected as it stems from the
fundamental symmetry of the theory, i.e. the foliation-preserving
diffeomorphism. Essentially, the lapse function $N$ represents a gauge
degree of freedom associated with space-independent time
reparametrization. Thus, it is very natural to restrict $N$ to be
independent of spatial coordinates. This is the projectability
condition. Of course, a space-independent $N$ cannot be transformed to
a space-dependent function by foliation-preserving diffeomorphism. The 
projectability condition then implies that the Hamiltonian constraint is
not a local equation satisfied at each spatial point but an equation
integrated over a whole space. This point was already made clear by
Ho\v{r}ava in \cite{Horava:2009uw} (See footnote
\ref{footnote:projectability} of the present paper.).

Abandoning the projectability condition and imposing a local version of 
the Hamiltonian constraint would result in phenomenological 
obstacles~\cite{Charmousis:2009tc} and theoretical
inconsistencies~\cite{Li:2009bg}. Note that a strong self-coupling of
the scalar graviton reported in \cite{Charmousis:2009tc} is
not a problem if there is a phenomenon analogous to Vainshtein
effect~\cite{Vainshtein:1972sx} since, unlike massive
gravity~\cite{Fierz:1939ix}~\footnote{In the case of massive gravity,
Vainshtein effect~\cite{Vainshtein:1972sx} removes the vDVZ
discontinuity~\cite{vanDam:1970vg,Zakharov:1970cc} but simultaneously
makes quantum corrections uncontrollable~\cite{ArkaniHamed:2002sp}. For
this reason, the massive gravity theory without a UV completion does not
have predictability even at macroscopic scales.}, Ho\v{r}ava-Lifshitz
gravity is supposed to be UV complete. Other problems reported in
\cite{Charmousis:2009tc,Li:2009bg} disappear if the projectability
condition is respected and if only the global Hamiltonian constraint is
imposed. The Poisson brackets of constraints form a closed structure 
since there is only one Hamiltonian constraint and it commutes with
itself~\cite{Horava:2008ih}. The divergent coupling of the scalar
graviton to matter source does not exist in the absence of local
Hamiltonian constraint~\footnote{If the local Hamiltonian constraint
were not used, then eq.~(68) of \cite{Charmousis:2009tc} would not show
a divergent coupling. This can be seen by moving the first term (written
in term of $\dot{\sigma}$) in the r.h.s. to the l.h.s.}. In conclusion,
both theoretical consistencies and phenomenological viability require
that the Hamiltonian constraint is not a local equation but an equation
integrated over a whole space.

The global Hamiltonian constraint is less restrictive than its local
version, and allows a richer set of solutions than in general
relativity. We have shown that a component which behaves like
pressureless dust emerges as an ``integration constant'' of dynamical
equations and momentum constraint equations. Consequently, classical
solutions to the infrared limit of Ho\v{r}ava-Lifshitz gravity can mimic
general relativity plus cold dark matter. The ``dark matter'' satisfies
the (non-)conservation equation (\ref{eqn:DMconservation}), whose source
term is turned on if a part of $4$-dimensional diffeomorphism invariance
is broken by either the (real) matter sector or the gravity
sector. In the IR, the source term vanishes and the standard
conservation equation (\ref{eqn:DMconservationIR}) holds. Also, the
``modified Einstein equation'' (\ref{eqn:modEin-general}) leads to the
Poisson equation (in a gauge with $N=N(t)$). Of course the ``dark
matter'' can cluster.

\section*{Note added}

After the present paper had appeared, the emergence of the ``dark
matter'' was confirmed in several
papers~\cite{Kocharyan:2009te,Blas:2009yd,Kobakhidze:2009zr}. For 
example, eq.~(22) of \cite{Kobakhidze:2009zr} is exactly equivalent to
Einstein gravity plus cold dark matter. Obviously, it is not appropriate 
to interprete it as a scalar-tensor theory. (This is obvious if we notice that we
usually do not try to interprete other cold dark matter models as
scalar-tensor theories.)  Nonetheless, ref.~\cite{Kobakhidze:2009zr}
did. Also, the analysis of ref.~\cite{Kobakhidze:2009zr} is not capable
of seeing discontinuity in the limit $\lambda\to 1$ since $\lambda$ is
set to $1$ from the beginning.

Ref.~\cite{Blas:2009yd} has three statements about Ho\v{r}ava-Lifshitz
gravity with the projectability condition: (i) "dark matter" forms
caustics; (ii) "dark matter" is described by ghost
condensate~\cite{ArkaniHamed:2003uy}; (iii) the scalar sector gets
strongly coupled at the scale $\Lambda\sim\rho_d^{1/4}$ even with 
$\lambda\ne 1$. Actually, these three comments are not correct for the 
following reasons. (i) They did not take into account repulsive gravity
due to nonlinear higher curvature terms (see the second-to-the-last
paragraph in Sec.~\ref{sec:CDM} of the present paper and also
ref.~\cite{Mukohyama:2009tp}). (ii) Ghost condensate and 
Ho\v{r}ava-Lifshitz gravity have different symmetries. In particular,
unlike ghost condensate, the invariance under (space-independent) time 
reparametrization forbids $h_{00}^2$ and thus $\dot{\pi}^2$. 
(iii) The strong coupling away from $\lambda=1$ found in
\cite{Blas:2009yd} indicates breakdown of their description, i.e. the
way dynamical degrees of freedom are identified, but does not imply 
inconsistency of the underlining UV theory. Note that the scalar
graviton does not get strongly coupled away from $\lambda=1$ as is clear
from e.g. eq.~(54) of ref.~\cite{Horava:2009uw}.  See also the last
paragraph in Sec.~\ref{sec:CDM} of the present paper. When we take the
limit $\lambda\to 1$, we have to take into account nonlinear
interactions carefully to see if there is an analogue of the Vainshtein
effect~\cite{Vainshtein:1972sx}.

More recently, ref.~\cite{Bogdanos:2009uj} rediscovered ghost
instability in the regime $1/3<\lambda<1$. This can be seen already in
eq.~(54) of \cite{Horava:2009uw} and just implies that we need to
consider the region $\lambda >1$.

\section*{Acknowledgements}

The author would like to thank Andrei Frolov, Keisuke Izumi, Kazuya
Koyama, Miao Li, Hitoshi Murayama, Antonio Padilla, Oriol Pujolas,
Keitaro Takahashi and Takahiro Tanaka for useful discussions. The work
of the author was supported in part by MEXT through a Grant-in-Aid for
Young Scientists (B) No.~17740134, by JSPS through a Grant-in-Aid for
Creative Scientific Research No.~19GS0219, and by the Mitsubishi
Foundation. This work was supported by World Premier International
Research Center Initiative (WPI Initiative), MEXT, Japan. 


\end{document}